\documentclass[twocolumn]{aastex61}
\usepackage{epsfig}
\usepackage{epstopdf}
\usepackage{color}
\usepackage{amsmath}
\usepackage{makecell}
\usepackage{gensymb}
\usepackage{lipsum}
\usepackage{subfigure}
\usepackage{multirow}

\submitjournal{ApJ Letters}

\shorttitle{Helium in HAT-P-11b}
\shortauthors{Mansfield et al.}

\begin{document}

\title{Detection of Helium in the Atmosphere of the Exo-Neptune HAT-P-11b}

\correspondingauthor{Megan Mansfield}
\email{meganmansfield@uchicago.edu}

\author{Megan Mansfield}
\affiliation{Department of Geophysical Sciences, University of Chicago, 5734 S. Ellis Avenue, Chicago, IL 60637, USA}

\author{Jacob L.\ Bean}
\affiliation{Department of Astronomy \& Astrophysics, University of Chicago, 5640 S. Ellis Avenue, Chicago, IL 60637, USA}

\author{Antonija Oklop\v ci\'c}
\affiliation{Harvard-Smithsonian Center for Astrophysics, Harvard University, Cambridge, MA 02138, USA}

\author{Laura Kreidberg}
\affiliation{Harvard-Smithsonian Center for Astrophysics, Harvard University, Cambridge, MA 02138, USA}

\author{Jean-Michel D\'esert}
\affiliation{Anton Pannekoek Institute for Astronomy, University of Amsterdam, 1090 GE Amsterdam, Netherlands}

\author{Eliza M.-R. Kempton}
\affiliation{Department of Astronomy, University of Maryland, College Park, MD 20742, USA}
\affiliation{Department of Physics, Grinnell College, Grinnell, IA, 50112, USA}

\author{Michael R. Line}
\affiliation{School of Earth and Space Exploration, Arizona State University, Tempe, AZ 85281, USA}

\author{Jonathan J. Fortney}
\affiliation{Department of Astronomy and Astrophysics, University of California, Santa Cruz, CA 95064, USA}

\author{Gregory W. Henry}
\affiliation{Center of Excellence in Information Systems, Tennessee State University, Nashville, TN 37209, USA}

\author{Matthias Mallonn}
\affiliation{Leibniz-Institut f\"ur Astrophysik Potsdam, 14482 Potsdam, Germany}

\author{Kevin B. Stevenson}
\affiliation{Space Telescope Science Institute, Baltimore, MD 21218, USA}

\author{Diana Dragomir}
\affiliation{Kavli Institute for Astrophysics and Space Research, Massachusetts Institute of Technology, Cambridge, MA 02139, USA}
\affiliation{Hubble Fellow}

\author{Romain Allart}
\affiliation{Observatoire Astronomique de l'Universit\'e de Gen\`eve, Universit\'e de Gen\`eve, 1290 Versoix, Switzerland}

\author{Vincent Bourrier}
\affiliation{Observatoire Astronomique de l'Universit\'e de Gen\`eve, Universit\'e de Gen\`eve, 1290 Versoix, Switzerland}

\begin{abstract}

The helium absorption triplet at a wavelength of 10,833\,\AA\ has been proposed as a way to probe the escaping atmospheres of exoplanets. Recently this feature was detected for the first time using \textit{Hubble Space Telescope} (\textit{HST}) WFC3 observations of the hot Jupiter WASP-107b. We use similar \textit{HST}/WFC3 observations to detect helium in the atmosphere of the hot Neptune HAT-P-11b at the $4\sigma$ confidence level. We compare our observations to a grid of 1D models of hydrodynamic escape to constrain the thermospheric temperatures and mass loss rate. We find that our data are best fit by models with high mass loss rates of $\dot{M} \approx 10^{9}$ -- $10^{11}$\,g\,s$^{-1}$. Although we do not detect the planetary wind directly, our data are consistent with the prediction that HAT-P-11b is experiencing hydrodynamic atmospheric escape. Nevertheless, the mass loss rate is low enough that the planet has only lost up to a few percent of its mass over its history, leaving its bulk composition largely unaffected. This matches the expectation from population statistics, which indicate that close-in planets with radii greater than 2\,R$_{\oplus}$ form and retain H/He-dominated atmospheres. We also confirm the independent detection of helium in HAT-P-11b obtained with the CARMENES instrument, making this the first exoplanet with the detection of the same signature of photoevaporation from both ground- and space-based facilities.

\end{abstract}


\section{Introduction}
\label{sec:intro}

Close-in exoplanets are expected to experience atmospheric escape that is driven by the absorption of the copious high energy radiation they receive from their host stars \citep{Lammer03,Lecavelier04}. Such photoevaporation likely sculpts the observed population of close-in exoplanets, dividing small planets into two categories - those with radii smaller than $1.5R_{\oplus}$, which are likely rocky cores stripped of any primordial light-element atmospheres, and those with radii larger than $2R_{\oplus}$, which retain some hydrogen and helium in their atmospheres \citep{Lopez2013,Owen2013,Owen2017,Fulton2017,VanEylen18}.

The theories describing the recently discovered planet radius gap can be refined through observations of escaping atmospheres, which will lead to a better understanding of the physics of photoevaporation. Atmospheric escape has been detected through observations of hydrogen absorption in the Lyman $\alpha$ line for two hot Jupiters \citep[HD\,209458b and HD\,189733b;][]{Vidal2003,Vidal2004,Lecavelier10} and two hot Neptunes \citep[GJ\,436b and GJ\,3470b;][]{Ehrenreich2015,bourrier18}. However, interstellar absorption limits these observations to nearby planets with gas escaping at high velocities.

Another possible signature of an escaping atmosphere that is less affected by interstellar absorption is the helium triplet at 10,833\,\AA\ \citep[note this is the wavelength of the feature in vacuum;][]{Seager2000,Oklopcic2018}. This feature was recently detected for the first time by \citet{Spake2018} using \textit{HST} observations of WASP-107b. In this paper we present similar \textit{HST} observations for the hot Neptune HAT-P-11b \citep{bakos10}. The original motivation for our program was to precisely determine the atmospheric water abundance of the planet and further constrain the exoplanet mass-metallicity relation \citep[e.g.,][]{Kreidberg2014b}. However, as was true for the \citet{Spake2018} observations, these data also presented the serendipitous opportunity to search for the previously-theorized but unexploited-until-recently He triplet. Our analysis of these data was further inspired by the presentation of R.\ Allart at the Exoplanets II conference in July 2018 showing a detection of the He feature in HAT-P-11b using ground-based data from CARMENES \citep{Allart2018}.

Our observations of HAT-P-11b have yielded the second detection of helium escaping from a planet using \textit{HST}, and a new detection of photoevaporation from a Neptune-sized exoplanet, and so help to constrain the nature of photoevaporation for planets smaller than Jupiter. Furthermore, HAT-P-11b now becomes the first exoplanet with the detection of the same signature of photoevaporation from both ground- and space-based facilities. In \S\ref{sec:obs} we describe our observations and data reduction process. In \S\ref{sec:analyze} we compare our observations to models of photoevaporation, and we summarize our findings in \S\ref{sec:discuss}.

\section{Observations and Data Reduction}
\label{sec:obs}

\subsection{{\textit HST} Data}
We observed five transits of HAT-P-11b between 14 September and 26 December 2016 using the \textit{HST} WFC3 IR detector as part of program GO-14793. We used the G102 grism to measure the transmission spectrum of HAT-P-11b between 0.8 -- 1.15\,$\mu$m. Each visit consisted of four consecutive \textit{HST} orbits in which HAT-P-11 was visible for approximately 56 minutes per orbit. At the beginning of each orbit, we took a direct image of the target with the F126N narrow-band filter for wavelength calibration.

The observations were taken in the spatial scan mode with the 256 x 256 subarray using the SPARS10, NSAMP=12 readout pattern, resulting in an exposure time of 81.089\,s. We used a scan rate of 0.25\,arcsec/s, which produced spectra extending approximately 180 pixels in the spatial direction and with peak pixel counts of about 45,000 electrons per pixel. We used bi-directional scans, which yielded 25 exposures per orbit and a duty cycle of 74\%.

We reduced the \textit{HST} data using the data reduction pipeline described in \citet{Kreidberg2014}, with the addition of an extra step to subtract light from a background star that overlapped with the spectrum of HAT-P-11. The spectrum of the background star was measured in one visit in which it was separated from HAT-P-11, and then subtracted out of each image, accounting for the fact that its position along both the dispersion and spatial axes changed between visits. This subtraction did not substantially change the final shape of the spectrum.

\begin{figure*}
    \centering
    \includegraphics[width=\linewidth]{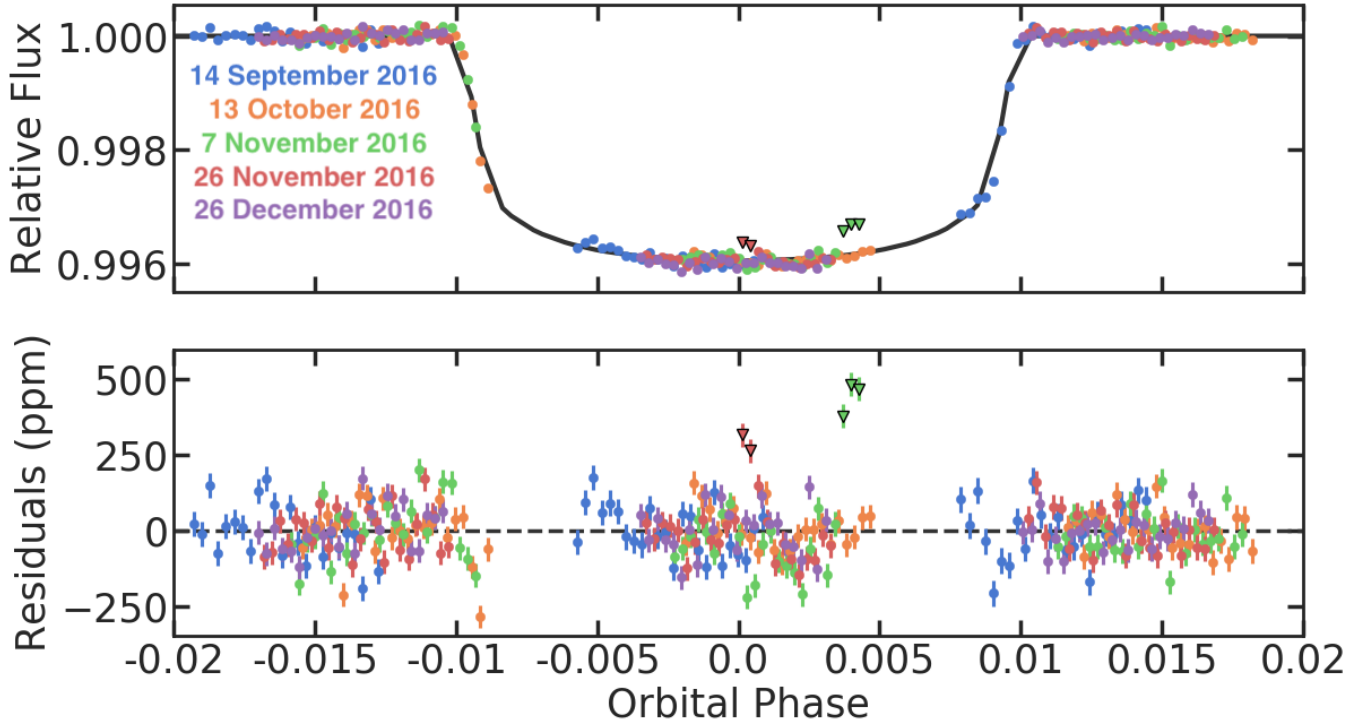}
    \vspace{-23 mm}
    \caption{Best fit broadband white light curve (top) and residuals to the fit (bottom). The two sets of triangular points outlined in black, which have residuals greater than 250\,ppm, are the starspot crossings that were removed before the data analysis. The fit had $\chi^{2}_{\nu}=4.67$ and an average residual of 75\,ppm.}
    \label{fig:whitelight}
\end{figure*}

Following standard procedure for \textit{HST} transit observations, we discard the first orbit of each visit. We also discard the first exposure from each orbit to improve the quality of the fit. Additionally, three points in the third visit and four points in the fourth visit were removed because they showed higher relative fluxes consistent with starspot crossings, as can be seen in Figure \ref{fig:whitelight}.

We used the WFC3 G102 wavelength calibration outlined in \citet{Kuntschner2009} to determine the wavelength of each pixel in our spectra from the direct image positions. However, we found that the calibrated data appeared to be shifted in wavelength space compared to the expected grism throughput. Both the red and blue edges of the spectrum were shifted by the same amount, but the shift was different for each visit, varying between 6 and 18\,\AA. This is likely due to the fact that \textit{HST} must move slightly after taking the direct image in order to correctly position the spatial scan so it fits on the detector. The positioning of the spectral images relative to the direct image has an uncertainty of roughly 12\,\AA\ \citep{WFC3Handbook}. We shifted the wavelengths to the correct values using a least-squares fit to the drop-offs at both the red and blue edges of the throughput.

We made a broadband spectrum by binning the observations into 16-pixel channels, resulting in 12 light curves with resolution $R\approx40$. To search for the narrow helium feature, we summed the region of the spectrum between 10,590\,\AA\ and 11,150\,\AA\ into 22 overlapping bins that were 2 pixels wide ($\approx49$\,\AA, for comparison a resolution element is $\approx67$\,\AA). Each bin was shifted by one pixel from the previous bin. We also summed over the entire 0.85 -- 1.13\,$\mu$m wavelength range to make a white light curve.

We fit the white light curve with the model described in \citet{Kreidberg2014}, which includes a transit model \citep{Kreidberg2015b} and a systematics model based on \citet{Berta2012}. The orbital period and eccentricity were fixed to $P=4.8878024$\,days and $e=0.265$, respectively \citep{Huber2017}. We placed Gaussian priors on $a/R_{s}$ and $i$ with mean and standard deviations of $17.125 \pm 0.060$ and $89.549 \pm 0.114$, respectively \citep{Fraine2014}. We fixed limb darkening coefficients to theoretical quadratic limb darkening models predicted from ATLAS models \citep{Espinoza2015, Castelli2004}. Including the instrument systematics, the fit to the white light curve contained a total of 24 free parameters (a normalization constant, visit-long linear slope, scaling factor to correct for an offset between scan directions, and the planet-to-star radius ratio varied between visits).

We estimated the parameters with a Markov Chain Monte Carlo (MCMC) fit using the {\fontfamily{bch}\selectfont emcee} package for Python \citep{Foreman2013}. Figure \ref{fig:whitelight} shows the best-fit broadband white light curve. The best-fit broadband white light curve had $\chi^{2}_{\nu}=4.67$ and residuals of 75\,ppm, which is typical for WFC3 observations of transiting planets orbiting bright host stars. The white light transit depth was $3371 \pm 15$\,ppm.

We fit the spectroscopic light curves using the divide-white technique of \citet{stevenson14} and \citet{Kreidberg2014}. For the narrowband spectroscopic light curves surrounding the helium feature, we used a region spanning 10,590 -- 11,150\,\AA\ to determine the white light curve systematics. The narrowband white light curve had $\chi^{2}_{\nu}=3.27$ and residuals of 138\,ppm. We fixed $a/R_{s}$ and $i$ to the prior mean values and the mid-transit times were fixed to the best-fit values from the white light curve fit. Before fitting each spectroscopic light curve, we rescaled the uncertainties by a constant factor such that each light curve had $\chi^{2}_{\nu}=1$ to give more conservative error estimates.

\begin{figure}
    \centering
    \includegraphics[width=\linewidth]{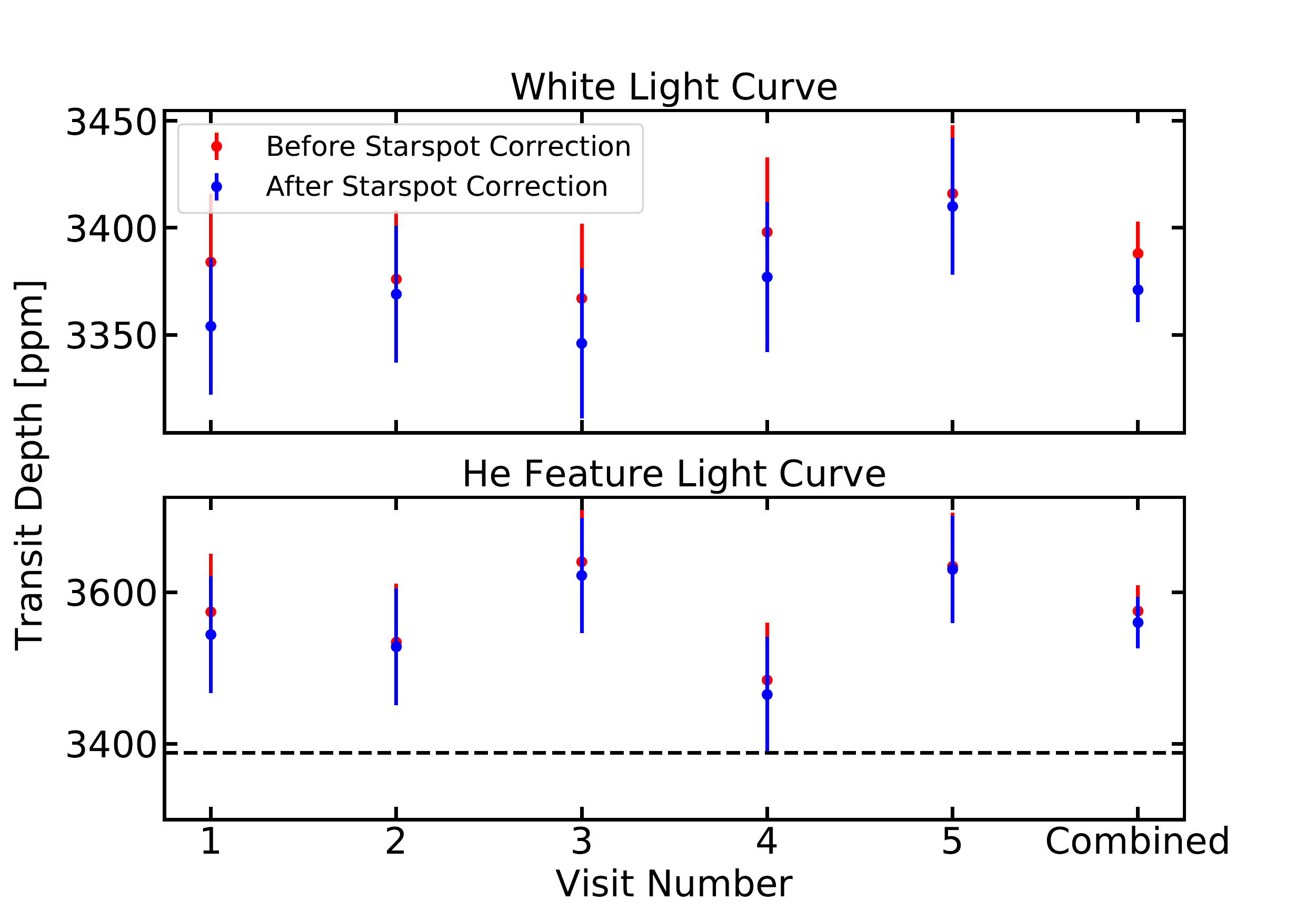}
    \caption{Individual white light curve transit depths (top) and transit depths in the narrow band containing the helium feature (bottom) for each visit. The final point shows the combined measurement from all five visits. For comparison, the dashed line in the lower plot shows the broadband white light transit depth from combining all five visits. Red and blue points show values before and after the starspot correction, respectively. For both the white light curve and the helium feature, the transit depths across all five visits were within $2\sigma$ of each other.}
    \label{fig:visitcomp}
\end{figure}

\begin{figure*}
    \centering
    \includegraphics[width=0.9\linewidth]{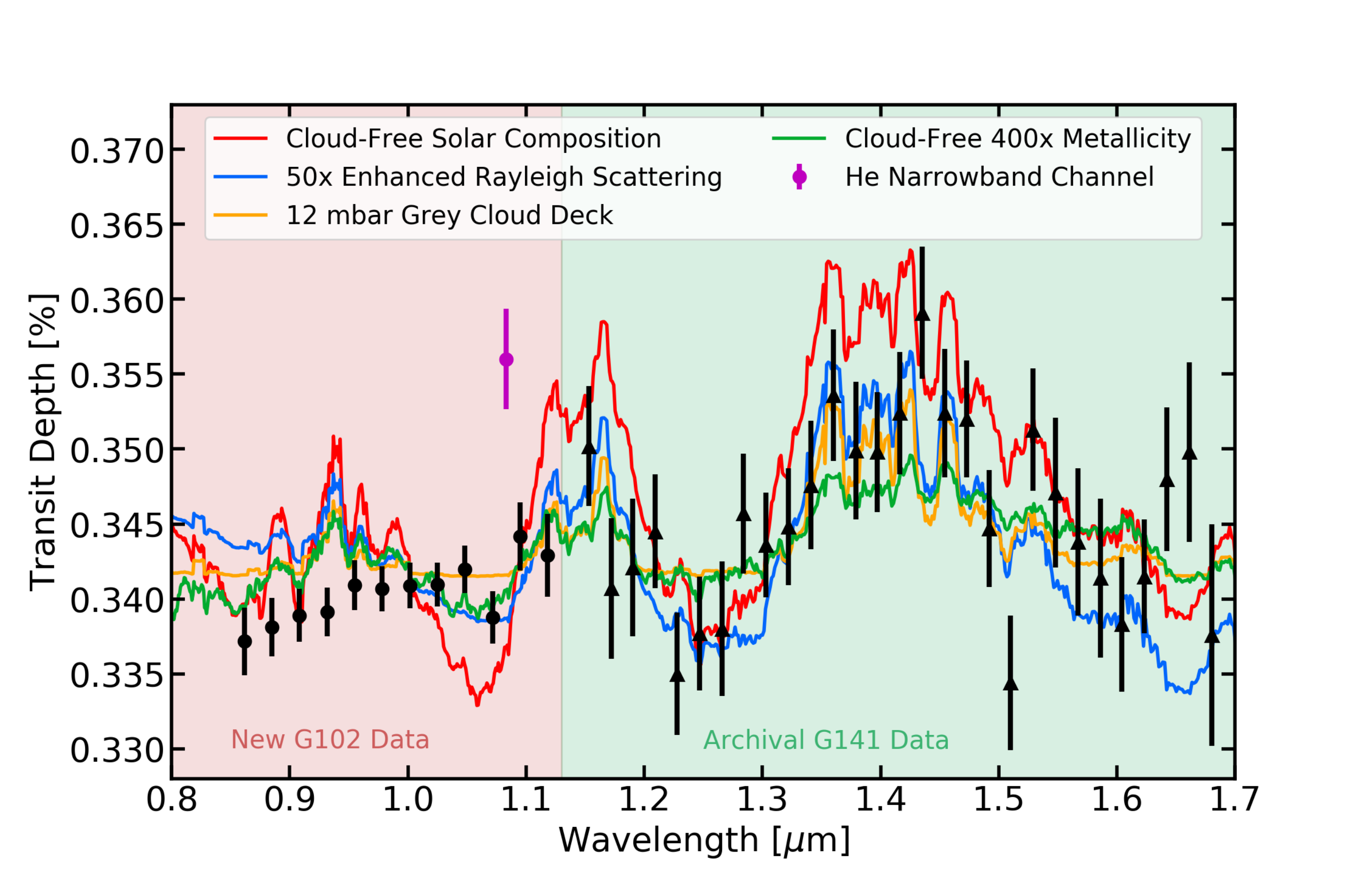}
    \caption{Spectrum of HAT-P-11b (black points with $1\sigma$ error bars) compared to a variety of models (lines) with different cloud parameterizations, which are described in \S~\ref{sec:analyzebroad}. Circular points are our G102 data (in bins $\approx233$\,\AA\ wide), and triangular points are G141 data from \citet{Fraine2014}. 
    The magenta point shows the transit depth in the 49-\AA\ wide channel containing the helium feature at 10,833\,\AA. The grey cloud deck model fits best, but does not match the upward slope seen between $0.8-1.1$~$\mu$m.}
    \label{fig:broadmodels}
\end{figure*}

\begin{figure*}
\centering
\includegraphics[width=\linewidth]{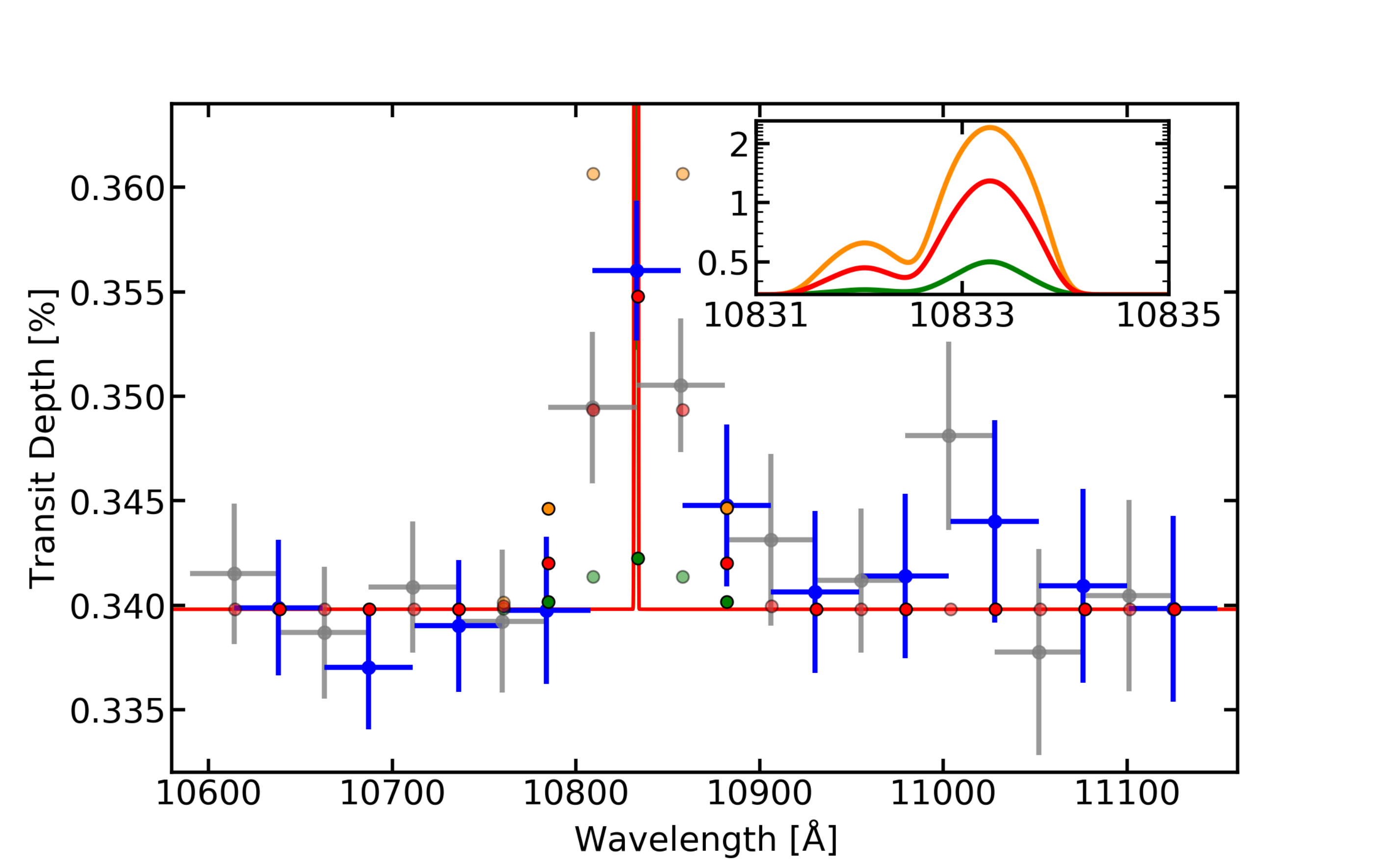}
\caption{\label{fig:narrow} Narrowband spectrum of HAT-P-11b (blue and grey points with 1$\sigma$ error bars) compared to three 1D models of hydrodynamic escape. The red line shows a model with $T=7,000$~K and a mass loss rate of $2.5\times 10^{10}$~g\,s$^{-1}$, which provides an excellent match to the data. For comparison, the green and orange lines show models with $T=7,000$~K and mass loss rates of $6.3\times 10^{9}$~g\,s$^{-1}$ and $5.0\times 10^{10}$~g\,s$^{-1}$, respectively. These models are inconsistent with the the data at $\geq3\sigma$ confidence. Blue points show non-overlapping bins. Red, green, and orange points outlined in black show the models convolved with the G102 instrument resolution \citep{Kuntschner2009} and binned to the sampling of the data. The inset shows the models at high resolution.}
\end{figure*}

Figure \ref{fig:visitcomp} shows the individual transit depths for each visit for both the broadband white light curve and the narrow band containing the helium feature. In both cases, the individual transit depths are all within $2\sigma$ of each other. Figure \ref{fig:broadmodels} shows the broadband spectrum, Figure \ref{fig:narrow} shows the narrowband spectrum, and Table \ref{tab:data} lists the transit depth in each channel. The derived transmission spectrum is largely featureless with the exception of a deeper transit in the narrowband spectroscopic channel corresponding to the unresolved infrared He triplet, which deviates from the surrounding continuum at the 4$\sigma$ confidence level.

\begin{table}
\centering
\caption{Broadband (left) and narrowband (right) transmission spectra of HAT-P-11b.}
\label{tab:data}
\begin{tabular}{>{\centering \arraybackslash}p{2.0 cm} | >{\centering \arraybackslash}p{1.5 cm} || >{\centering \arraybackslash}p{2.0 cm} | >{\centering \arraybackslash}p{1.5 cm}}
\hline \hline
Wavelength (\AA) & Transit Depth (ppm) & Wavelength (\AA) & Transit Depth (ppm) \\
\hline
8500 -- 8733 & $3371\pm23$ & 10614 -- 10663 & $3399\pm32$ \\
8733 -- 8967 & $3381\pm19$ & 10663 -- 10712 & $3370\pm30$  \\
8967 -- 9200 & $3389\pm18$ & 10712 -- 10760 & $3390\pm32$ \\
9200 -- 9433 & $3391\pm16$ & 10760 -- 10809 & $3398\pm35$ \\
9433 -- 9667 & $3409\pm17$ & 10809 -- 10858 & $3560\pm34$ \\
9667 -- 9900 & $3406\pm15$ & 10858 -- 10907 & $3448\pm39$ \\
9900 -- 10133 & $3409\pm15$ & 10907 -- 10955 & $3406\pm39$ \\
10133 -- 10367 & $3410\pm15$ & 10955 -- 11004 & $3414\pm39$ \\
10367 -- 10600 & $3420\pm16$ & 11004 -- 11053 & $3440\pm48$ \\
10600 -- 10833 & $3388\pm17$ & 11053 -- 11101 & $3409\pm46$ \\
10833 -- 11067 & $3442\pm23$ & 11101 -- 11150 & $3398\pm45$ \\
11067 -- 11300 & $3429\pm28$ & &  \\
\hline \hline
\end{tabular}
\end{table}

\subsection{Ground-based Photometry}
We obtained ground-based photometric monitoring from two sources to correct the transmission spectrum for changes in stellar activity. We obtained multi-color photometric monitoring with the 1.2\,m robotic STELLA telescope using its wide-field imager WiFSIP \citep{Strassmeier2004}. Data were taken from June -- October 2016 and March -- July 2017 in Johnson B and V in nightly blocks of four exposures per filter, which were averaged during data reduction. The data reduction followed the procedure of \citet{Mallonn2018}. We also monitored the photometric variability using the Tennessee State University Celestron 14 inch (C14) automated imaging telescope (AIT), which is located at Fairborn Observatory \citep{Henry1999}. These observations consisted of 89 nights between 27 September 2015 and 7 January 2017.

We interpolated between these photometric observations to estimate the starspot covering fraction at the time of each \textit{HST} observation. We used the relative amplitude of two occulted starspots in the white light curve to determine the temperature difference between the starspots and the star, which we found to be approximately 150\,K. We assumed that the starspots were circular and that the entire starspot was occulted by the planet each time. We applied a wavelength-dependent correction to the $R_{p}/R_{s}$ values measured from the WFC3 data to account for variations in starspot coverage between visits, assuming that both the spotted and unspotted parts of the star emitted as blackbodies and assuming that the mean photometric magnitude occurred when the spot covering fraction was 3\%, which is the mean spot coverage for HAT-P-11b \citep{Morris2017a}.

We followed the method of \citet{Spake2018} to check whether the He absorption feature we observed could be caused by stellar activity. They found that the He triplet equivalent width for a K4-type star like HAT-P-11 should be no larger than $\approx0.4$\,\AA, according to both theoretical calculations and observations of K-dwarf stars. For our 49\,\AA-wide bins, this corresponds to a change in the transit depth of 30~ppm. The helium feature we observe has a depth of $162\pm36$\,ppm, which is more than $3.5\sigma$ larger than the expected depth due to stellar activity. Our argument that the He feature can not be due to stellar activity alone is strengthened by the fact that our observations occurred near a minimum in HAT-P-11's activity cycle \citep{Morris2017b}.

\section{Analysis}
\label{sec:analyze}

\subsection{Broadband Spectrum}
\label{sec:analyzebroad}
We computed transmission spectra models to compare with the WFC3+G141 data from \citet{Fraine2014} and our new broadband G102 data using the  \texttt{Exo-Transmit} code \citep{kempton17}. We tested models with a cloud-free, solar composition; enhanced metallicity; a grey cloud deck; and enhanced Rayleigh scattering. Figure \ref{fig:broadmodels} shows these different cloud parameterizations compared to the spectrum of HAT-P-11b. The best-fitting model had grey clouds at 12~mbar and $\chi^{2}_{\nu}=1.38$. The models with a cloud-free solar composition, 50$\times$ enhanced Rayleigh scattering, and 400$\times$ enhanced metallicity had $\chi^{2}_{\nu}=4.22$, $2.49$, and $1.53$, respectively. However, none of the models we tested produced a particularly good fit to both the archival G141 data and our new G102 data, as none of them produced the upward slope seen between $0.8-1.1$~$\,\mu$m while simultaneously matching the water feature seen around 1.4\,$\mu$m. One possible explanation for the upward slope we observe is the influence of unocculted, bright regions on the star \citep{Rackham2018}. We leave a more detailed study of the broadband transmission spectrum for future work (Chachan et al.\ in prep) and turn our attention below to a detailed study of the narrowband spectrum around the He triplet feature.

\subsection{Narrowband Helium Spectrum}
\label{sec:analyzenarrow}
We compared the helium absorption feature detected in our \textit{HST} data to a grid of models of hydrodynamic escape computed using the methods of \citet{Oklopcic2018}. Spherically-symmetric model atmospheres were constructed from 1D density and velocity profiles based on the isothermal 
Parker wind model \citep{Parker1958,Lamers1999}. The atmospheres were assumed to be composed of hydrogen and helium atoms in 9:1 number ratio. Hydrogen and helium atomic level populations were computed taking into account photoionization, recombination, and collisional transitions. To calculate the photoionization rates, we constructed a spectrum appropriate for HAT-P-11 (a K4 star) by averaging the observed spectra of K2 and K6 stars, obtained from the MUSCLES survey \citep{France2016}. With the planet mass, radius, and atmospheric composition fixed, and the mean molecular weight of the atmosphere evaluated iteratively (taking into account the free electrons produced by hydrogen photoionization), the remaining free parameters in the model are temperature and mass loss rate. 

We model the atmosphere of HAT-P-11b as being in the hydrodynamic escape regime because the planet has a low gravitational potential \citep{salz16a} and the Jeans parameter at the exobase \citep[$R>10\,R_{p}$;][]{salz16b} is $\lambda<1.3$. \citet{salz16b} also predict that the exobase should be above both the sonic point and the Roche radius for HAT-P-11b. Additionally, the Jeans escape rate is generally expected to be substantially lower than the hydrodynamic escape rate for a planet like HAT-P-11b with an unstable thermosphere \citep{Tian2005}.

We computed models for a range of thermospheric temperatures between $T$ = 3,000 -- 12,000\,K and total mass loss rates between $\dot{M}$ = $4.0 \times 10^{8}$ -- $2.5 \times 10^{11}$~g\,s$^{-1}$. We convolved each high-resolution model with a Gaussian function with the resolution of WFC3+G102, which is $R\approx 155$ at 10,400\,\AA\ \citep{Kuntschner2009}, before binning the models to the sampling of our observations. We then determined the $\chi^2$ of each model compared to the data using the three non-overlapping bins surrounding the helium feature that show a deviation from the baseline absorption level.

\begin{figure}
    \centering
    \includegraphics[width=\linewidth]{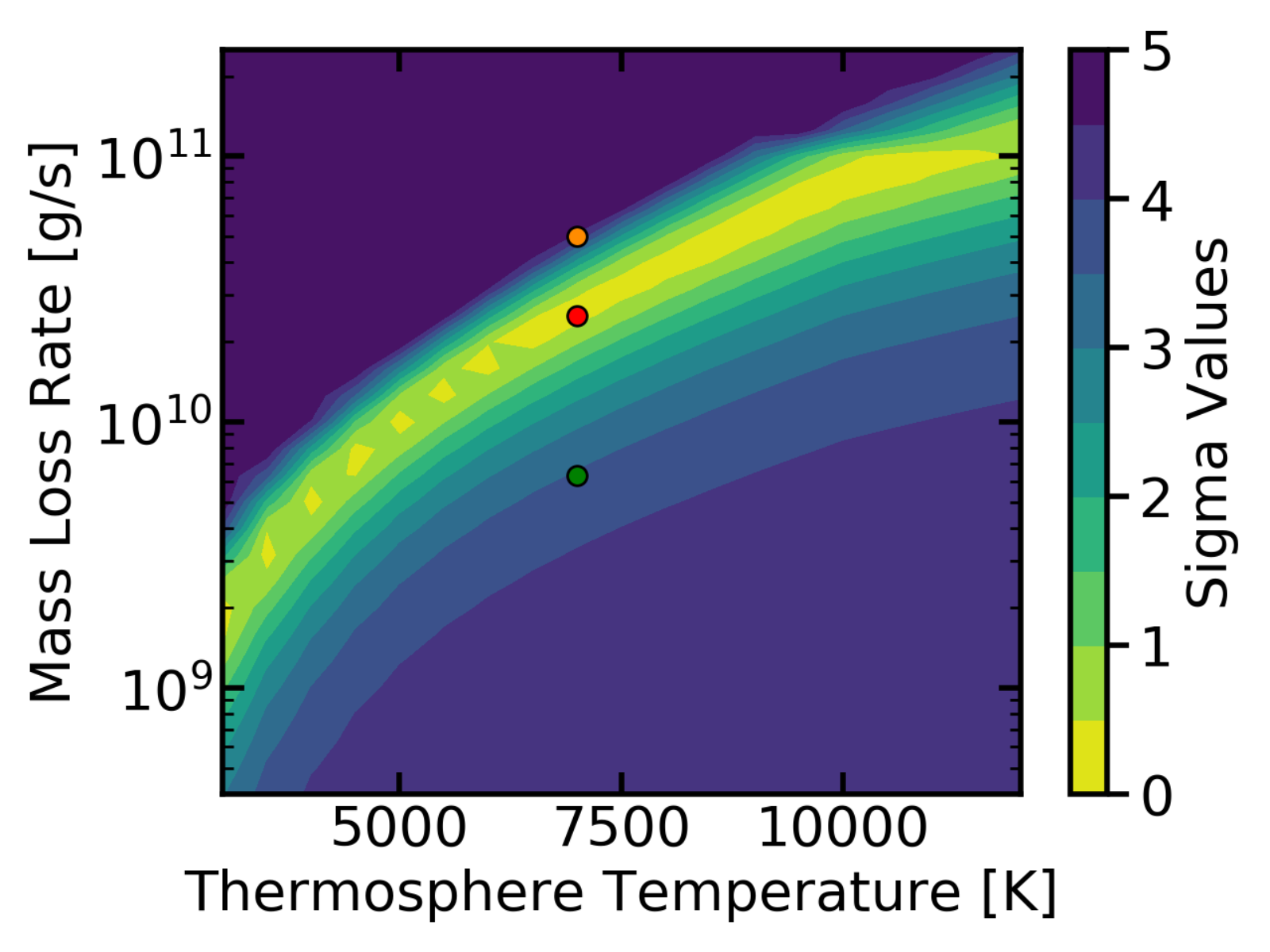}
    \caption{Contour plot showing the statistical significance of the deviation from a good fit for the comparison of the 1D grid model to our observations, as a function of the thermospheric temperature and mass loss rate. The red, green, and orange points outlined in black show the location in parameter space of the three models plotted in Figure \ref{fig:narrow}.
    }
    \label{fig:sig}
\end{figure}

Figure~\ref{fig:sig} shows a contour plot of the fit quality for comparing the grid models to the data cast in terms of the statistical significance of the deviation from a good fit. The banana-shaped contour of good fit quality indicates that the thermospheric temperature and the mass loss rate can be traded against each other to give an acceptable fit to the data for a wide range of values in each parameter. According to the mass continuity equation $\dot{M}=4\pi r \rho(r) v(r)$, the gas density required to produce the observed absorption signal can be achieved for different combinations of the mass loss rate and the gas radial velocity. For an isothermal Parker wind, the velocity is directly related to the gas temperature, thus giving rise to the mass loss rate - temperature degeneracy. This degeneracy can be reduced somewhat by resolving the shape of the absorption line. Even with the degeneracy between the temperature and the mass loss rate, a large portion of the parameter space can be excluded at high confidence despite the low-resolution of the \textit{HST} data due to the sensitivity of the He triplet to the gas density. Figure~\ref{fig:narrow} compares two example models from the excluded region of parameter space to the data, demonstrating the poor fit quality of such models.

The excess absorption in the He triplet that we detect suggests that HAT-P-11b's thermosphere extends to altitudes of at least 2\,R$_{p}$. \citet{salz16b} performed simulations of hydrodynamic escape from hot gas planets using a coupled photoionization and plasma simulation code with a general hydrodynamics code. For HAT-P-11b, \citet{salz16b} predict a temperature of $T\,\approx\,7,000$~K at 2\,R$_{p}$ and an overall escape rate of $\dot{M}\,=\,2 \times 10^{10}$~g\,s$^{-1}$. Despite the caveats given by \citet{salz16b} for their model (uncertainty in the stellar wind strength, magnetic effects, the effect of metals in the atmosphere, and the correction factor to account for dayside-only heating), Figure~\ref{fig:narrow} shows how their prediction falls within their estimated error of the region of good fit quality mapped in Figure~\ref{fig:sig}.

\begin{figure}
    \centering
    \includegraphics[width=\linewidth]{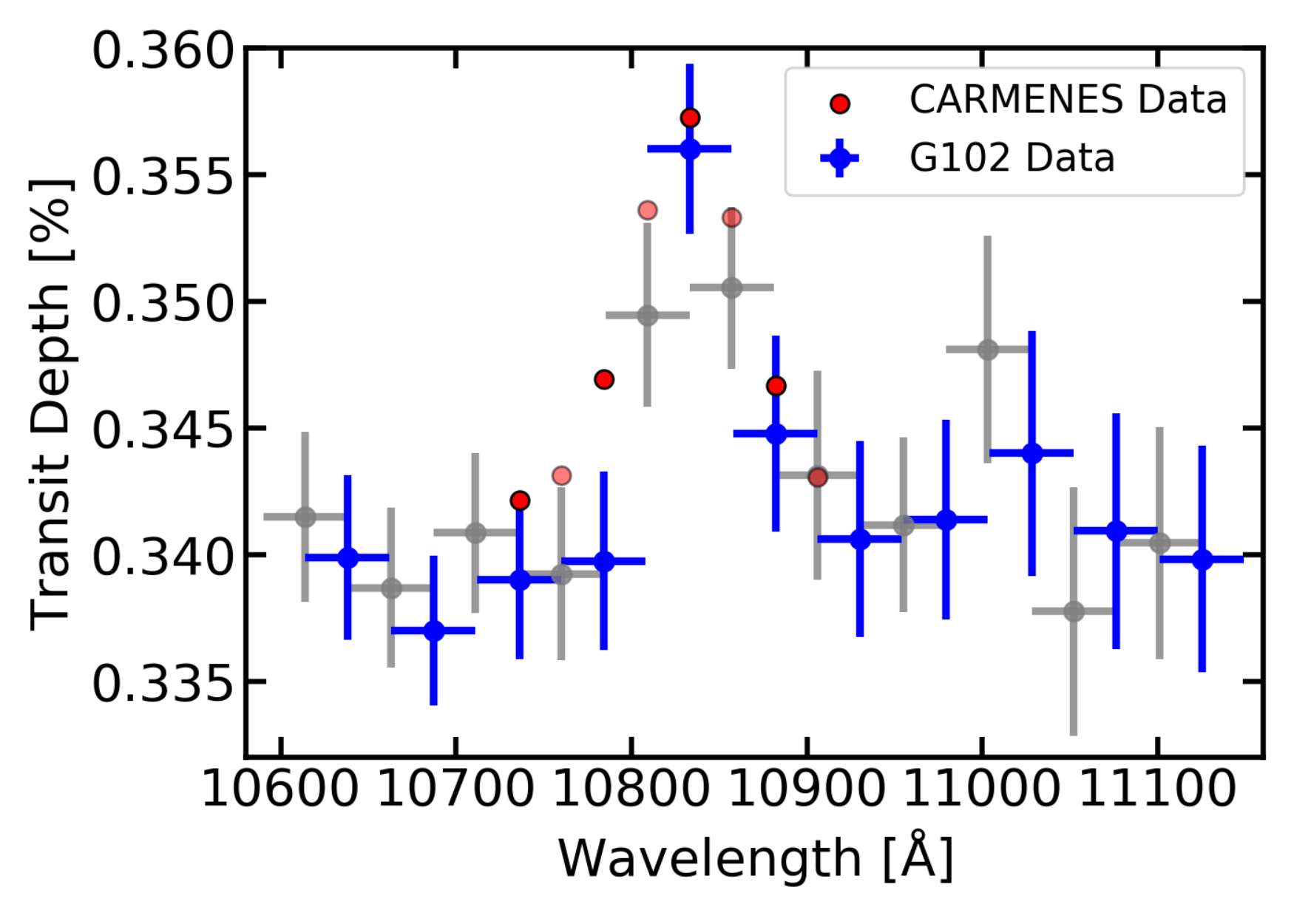}
    \caption{Narrowband G102 spectrum (blue and grey points) compared to recent observations of HAT-P-11b with the CARMENES spectrograph \citep{Allart2018}, convolved to the resolution of our data. The two data sets show excellent agreement in the size of the helium feature at 10,833\,\AA.}
    \label{fig:carmenes}
\end{figure}

We also compared our result to the data underlying the recent detection of the helium infrared triplet in HAT-P-11b using the CARMENES spectrograph \citep{Allart2018}. Figure \ref{fig:carmenes} shows the CARMENES spectrally resolved data convolved to the much lower resolution of our \textit{HST} data. Our data sets agree well on the size of the helium absorption feature. The confirmation of the ground-based data with our space-based observations, which are free from the influence of telluric lines and transparency variations, adds significant confidence to the detection of this subtle feature.

\section{Discussion}
\label{sec:discuss}

We find an overall mass loss rate of $0.04-2.3$\% of the total mass of HAT-P-11b per billion years. Additionally, \citet{salz16b} predict a mass loss of $\approx0.8$\% over the first 100~Myr after planet formation, when the young star would have caused a much higher escape rate. A total mass loss of 1\% would change the radius of HAT-P-11b by 0.4\%, or 108~km \citep{Seager2007}. As is expected for a Neptune-sized gas giant, this amount of photoevaporation will have a negligible effect on the composition of HAT-P-11b over its lifetime.

Although the mass loss from HAT-P-11b is negligible, smaller sub-Neptunes can be significantly impacted by photoevaporation over their lifetimes. Close-in planets smaller than $1.5R_{\oplus}$ may even have their entire primary atmospheres stripped away \citep{Lopez2013,Owen2013,Owen2017,Fulton2017,VanEylen18}. Observations of the helium triplet at 10,833\,\AA\ provides us with a new way to characterize atmospheric loss from a wider range of planets, which will help to refine models of photoevaporation. Future observations of smaller planets at this wavelength will also help us better constrain the exact nature of super-Earth/sub-Neptune planets and how their atmospheres have evolved over time.

\acknowledgements
We thank David Ehrenreich and Christophe Lovis for discussion of the CARMENES data. Support for program GO-14793 was provided by NASA through a grant from the Space Telescope Science Institute, which is operated by the Association of Universities for Research in Astronomy, Inc., under NASA contract NAS 5-26555. J.L.B.\ acknowledges support from the David and Lucile Packard Foundation. D.D.\ acknowledges support provided by NASA through Hubble Fellowship grant HSTHF2-51372.001-A awarded by the Space Telescope Science Institute, which is operated by the Association of Universities for Research in Astronomy, Inc., for NASA, under contract NAS5-26555. The work of R.A.\ and V.B.\ has been carried out in the framework of the National Centre for Competence in Research ``PlanetS'' supported by the Swiss National Science Foundation (SNSF). R.A.\ and V.B.\ acknowledge the financial support of the SNSF. This project has received funding from the European Research Council (ERC) under the European Union’s Horizon 2020 research and innovation programme (project FOUR ACES; grant agreement No.\ 724427). Based partly on data obtained with the STELLA robotic telescopes in Tenerife, an AIP facility jointly operated by AIP and IAC.

\begin{figure}
    \centering
    \includegraphics[width=\linewidth]{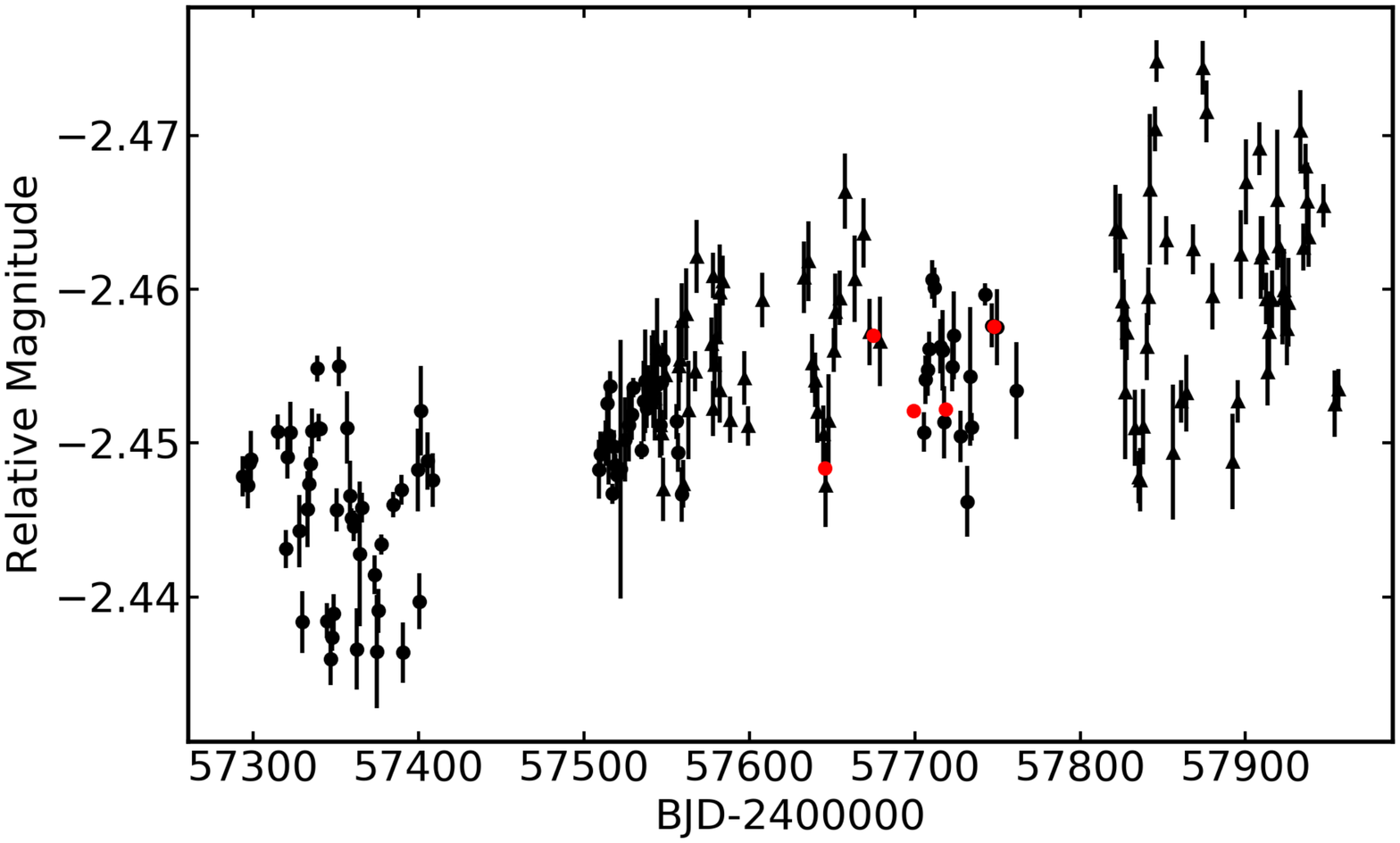}
    \caption{Ground-based photometry (black points) taken during the epoch of our \textit{HST} observations. Circular and triangular points show photometry taken with the Tennessee State University C14 AIT \citep{Henry1999} and the STELLA telescope \citep{Strassmeier2004}, respectively. The red points show the interpolated values at the times of our observations.}
    \label{fig:photometry}
\end{figure}

\begin{figure}
\centering
	\begin{subfigure}{}
	\includegraphics[width=0.75\linewidth]{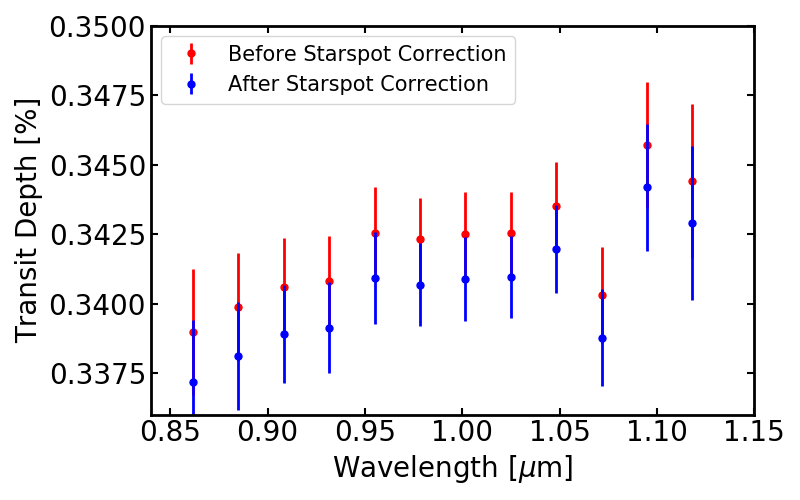}
	\end{subfigure}
	\begin{subfigure}{}
	\includegraphics[width=0.75\linewidth]{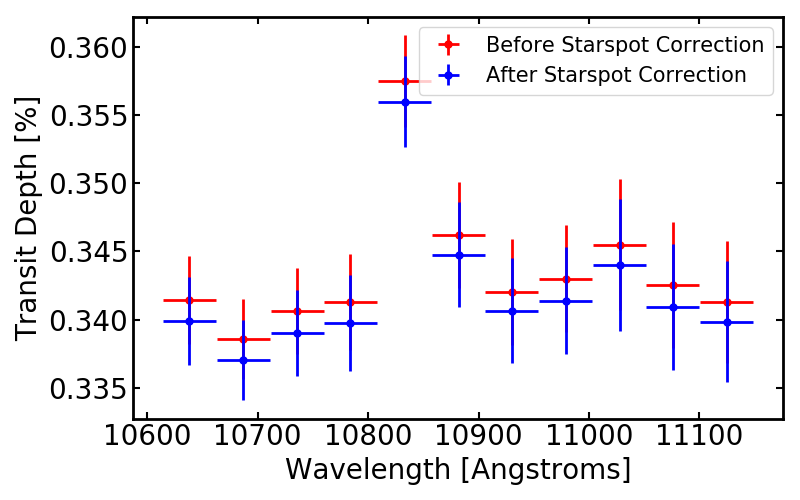}
	\end{subfigure}
\caption{\label{fig:starspotcorr} Effect of the correction for unocculted starspots on the broadband (top) and narrowband (bottom) spectra. The primary effect is to shift the entire spectrum up by about 20\,ppm, but in the broadband spectrum the correction also shifts the blue end of the spectrum up by slightly more than the red end of the spectrum.}
\end{figure}

\begin{figure}
\centering
	\begin{subfigure}{}
	\includegraphics[width=0.75\linewidth]{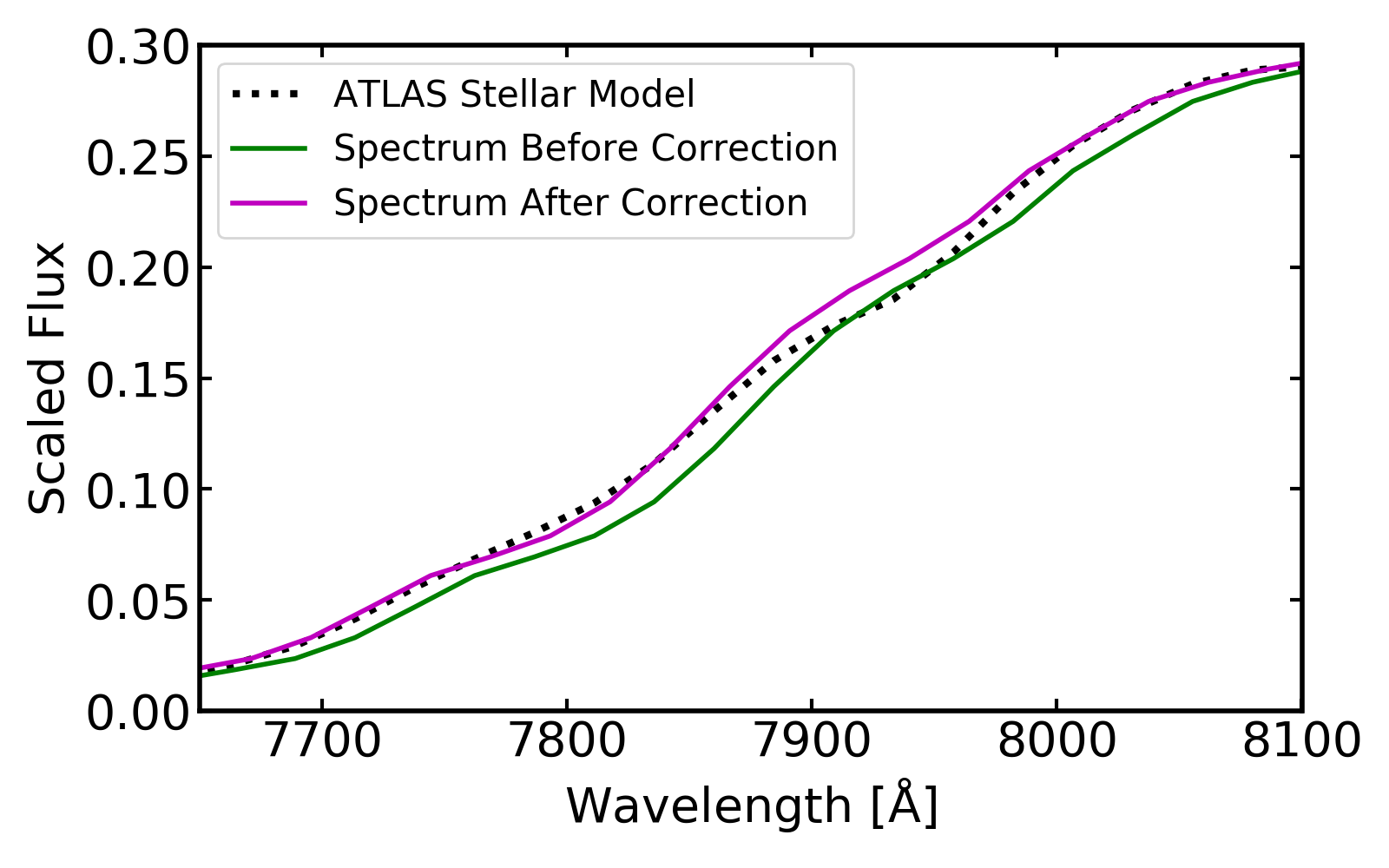}
	\end{subfigure}
	\begin{subfigure}{}
	\includegraphics[width=0.75\linewidth]{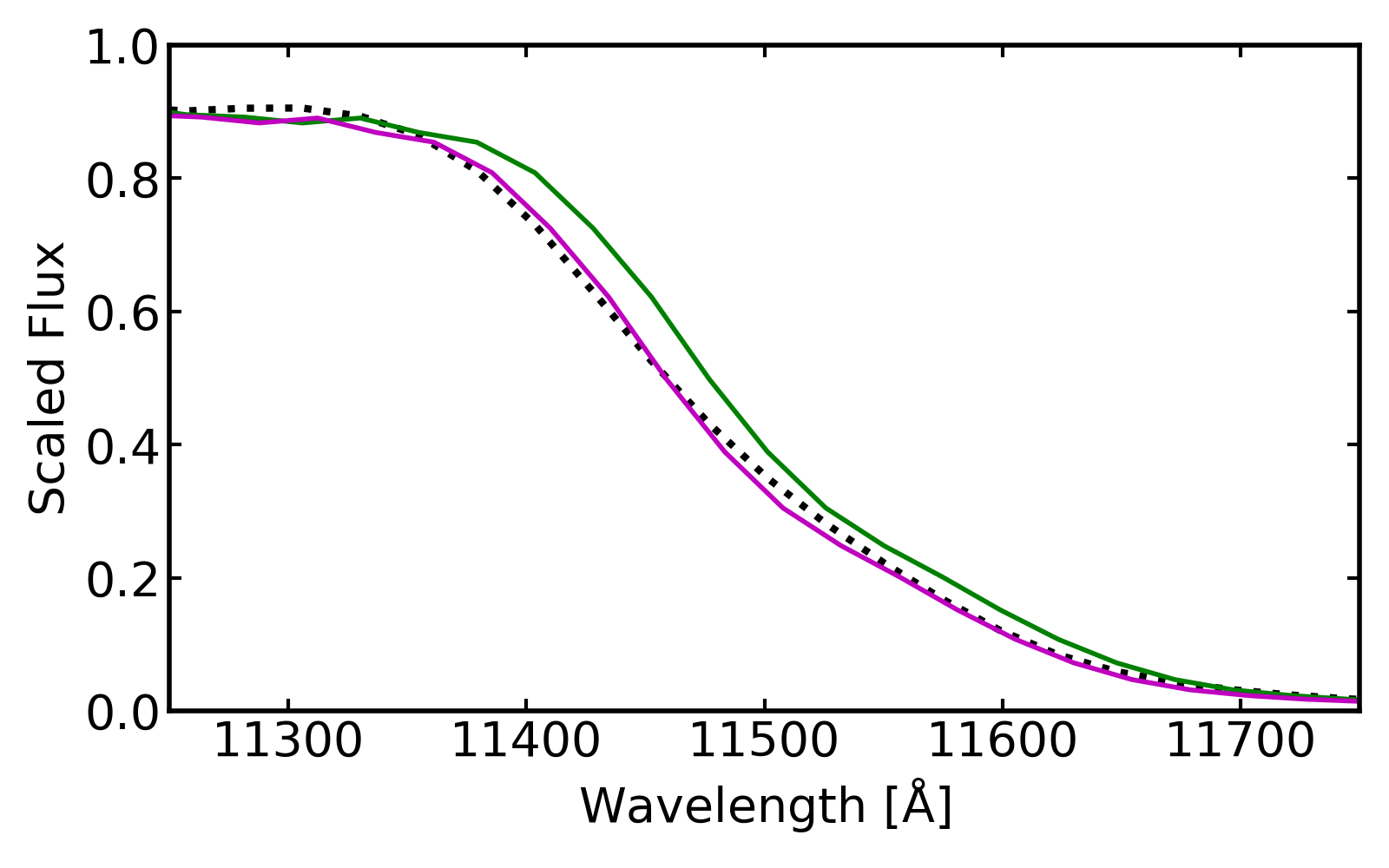}
	\end{subfigure}
\caption{\label{fig:wavecalib} Comparison between an ATLAS stellar model for HAT-P-11b combined with the G102 throughput (black dashed line) and the observed spectrum. The green line is the observed spectrum using the wavelength calibration outlined by \citet{Kuntschner2009} and assuming the distance between the direct image and the spectrum is exactly known, and the magenta line is the spectrum after adjusting the wavelength calibration. The plots show the blue (top) and red (bottom) edges of the spectrum, which were used to determine the wavelength shift.}
\end{figure}

\end{document}